\documentclass[runningheads]{llncs}
\usepackage{listings}
\usepackage{courier}
\usepackage{graphicx}
\usepackage[ruled,linesnumbered]{algorithm2e}
\usepackage{amsmath,amsfonts,amssymb}
\usepackage{todonotes}
\usepackage[hidelinks]{hyperref}
\usepackage{array, makecell}

\SetKwInput{KwGlobal}{Global}
\newcommand{\flag}{\mathit{flag}}

\newcommand{\wait}{\mathit{wait}}

\newcommand{\id}{\mathit{i}}
\newcommand{\node}{\mathit{n}}
\newcommand{\side}{\mathit{s}}
\newcommand{\Flag}{\mathit{Flag}}
\newcommand{\enter}{\mathit{enter}}
\newcommand{\leave}{\mathit{leave}}

\newcommand{\sibling}[1]{\mathit{sibling}_{#1}}
\newcommand{\siblingN}{\sibling{N}}
\newcommand{\initnode}[1]{\mathit{init\_node}_{#1}}
\newcommand{\initnodeN}{\initnode{N}}
\newcommand{\initside}{\mathit{init\_side}}
\newcommand{\nextnode}[1]{\mathit{next}_{#1}}
\newcommand{\nextN}{\nextnode{N}}
\newcommand{\overtakes}[1]{\mathit{overtakes}_{#1}}
\newcommand{\overtakesN}{\overtakes{N}}
\newcommand{\overtakeN}{\mathit{overtake}_{N}}

\usepackage[appendix=inline]{apxproof}
\theoremstyle{plain}
\newtheoremrep{theorem}{Theorem}
\newtheoremrep{lemma}{Lemma}
\newtheoremrep{proposition}{Proposition}
\renewcommand{\qedhere}{\qed}

\usepackage{apptools}
\AtAppendix{\counterwithin{lemma}{section}}

\usepackage{pifont}
\newcommand{\crossmark}{\ding{53}}%
\usepackage{booktabs}
\usepackage{multirow}

\usepackage{mathtools}
\DeclarePairedDelimiter{\ceil}{\lceil}{\rceil}
\DeclarePairedDelimiter{\floor}{\lfloor}{\rfloor}

\usepackage{floatflt}

\usepackage{tikz}
\usetikzlibrary{arrows,automata}
\usetikzlibrary{positioning}
\tikzset{initial text={},auto} 
\tikzstyle{every state}=[draw,shape=circle,inner sep=1pt,minimum size=12pt]

\lstdefinelanguage{mCRL2-inline}
{
	keywords={act,var,cons,end,eqn,glob,init,val,whr,sort,map,pbes,proc,struct},
	keywords=[2]{true,false,delta,tau},
	keywords=[3]{Bool,Nat,Real,Pos,Int,Set,Bag,List,Int2Nat,Pos2Nat,Int2Pos,min,max
	},
	keywords=[4]{hide,if,rename,sum,in,mu,nu,forall,exists,mod,allow,block,comm},
	keywords=[5]{nested,initial,state},
	numberstyle=\color{blue},
	comment=[l]\%,
	commentstyle=\slshape,
	keywordstyle=[1]\bfseries,
	keywordstyle=[2]\itshape,
	keywordstyle=[3]\itshape,
	keywordstyle=[4]\itshape,
	keywordstyle=[5]\bfseries\itshape,
	basicstyle=\ttfamily\footnotesize,
	flexiblecolumns=false,
	breaklines=true
}
[keywords,comments]
\lstdefinelanguage{mCRL2}
{
	keywords={act,var,cons,end,eqn,glob,init,val,whr,sort,map,pbes,proc,struct},
	keywords=[2]{true,false,delta,tau},
	keywords=[3]{Bool,Nat,Real,Pos,Int,Set,Bag,List,Int2Nat,Pos2Nat,Int2Pos,min,max
	},
	keywords=[4]{hide,if,rename,sum,in,mu,nu,forall,exists,mod,allow,block,comm},
	keywords=[5]{nested,initial,state},
	numberstyle=\color{blue},
	comment=[l]\%,
	commentstyle=\slshape,
	keywordstyle=[1]\bfseries,
	keywordstyle=[2]\itshape,
	keywordstyle=[3]\itshape,
	keywordstyle=[4]\itshape,
	keywordstyle=[5]\bfseries\itshape,
	basicstyle=\ttfamily\scriptsize,
	flexiblecolumns=false,
	breaklines=false
}
[keywords,comments]

\newcommand{\allow}[1]{\nabla_{\{#1\}}}
\newcommand{\comm}[1]{\Gamma_{\{#1\}}}

\newcommand{\false}{\mathit{false}}
\newcommand{\true}{\mathit{true}}

\title{Fair Mutual Exclusion for $N$ Processes (extended version)}
\author{Yousra Hafidi\orcidID{0000-0002-3543-6731} \and Jeroen J. A. Keiren\orcidID{0000-0002-5772-9527} \and Jan Friso Groote\orcidID{0000-0003-2196-6587}}

\institute{Eindhoven University of Technology, Eindhoven, The Netherlands
	\email{\{y.hafidi,j.j.a.keiren,j.f.groote\}@tue.nl}}

\begin{document}
	\maketitle

	\begin{abstract}
	Peterson's mutual exclusion algorithm for two processes has been generalized to $N$ processes in various ways.
	As far as we know, no such generalization is starvation free without making any fairness assumptions.
	In this paper, we study the generalization of Peterson's algorithm to $N$ processes using a tournament tree.
	Using the mCRL2 language and toolset we prove that it is not starvation free unless weak fairness assumptions are incorporated.
	Inspired by the counterexample for starvation freedom, we propose a fair $N$-process generalization of Peterson's algorithm.
	We use model checking to show that our new algorithm is correct for small $N$.
	For arbitrary $N$, model checking is infeasible due to the state space explosion problem, and instead, we present a general proof that, for $N \geq 4$, when a process requests access to the critical section, other processes can enter first at most $(N-1)(N-2)$ times.

	\end{abstract}

\section{Introduction}

Peterson's algorithm~\cite{Pet1981} is a classic, widely studied mutual exclusion algorithm for two processes. It satisfies all requirements desired from a mutual exclusion algorithm, e.g., mutual exclusion, starvation freedom, and bounded overtaking.

Peterson's algorithm can be generalized to $N$ processes using a tournament tree~\cite{PF1977}.
The tournament tree is a binary tree in which each of the leafs is assigned (at most) two processes that want to gain access to the critical section.
In each node of the tree, two processes compete using Peterson's algorithm for two processes. The winner moves up in the tournament tree.
Ultimately, the process that wins in the root of the tree gets access to the critical section.

This generalization is known to satisfy mutual exclusion.
However, the situation for starvation freedom is less clear.
Fokkink argues the algorithm satisfies starvation freedom~\cite{Fok2013}, but is not explicit about the assumptions needed for this.
Others argue that starvation freedom is not satisfied~\cite{Ray2013}, and that tournament tree solutions never satisfy bounded overtaking~\cite{Ala2005,Hes2017}.

To ensure starvation freedom, it turns out (weak) fairness assumptions are required.
In particular, if a process is in a state where it can continuously execute a particular statement, it will eventually be allowed to execute said statement.
To the best of our knowledge there is no generalization of Peterson's algorithm to $N$ processes that does not require fairness to ensure starvation freedom.

\paragraph{Contributions.}

We create an mCRL2 model~\cite{GM2014} of the generalization of Peterson's algorithm to $N$ processes using tournament trees, based on~\cite{Fok2013}.
We specify the most important requirements using the first order modal mu-calculus~\cite{GW2005}, and verify them using the mCRL2 toolset~\cite{BGK+2019} for instances with $N \leq 5$.
It is shown that the algorithm satisfies mutual exclusion. 
However, it is not starvation free. We present a counterexample illustrating that, when a process requests access to the critical section, a process in a different subtree can repeatedly request, and get, access to its critical section, preventing progress of the first process.

We propose a fair $N$-process mutual exclusion algorithm that introduces dependencies between the progress of all processes in order to rule out such counterexamples.
Using model checking we verify that this algorithm is starvation free. Furthermore, we verify that it satisfies bounded overtaking for $N \leq 5$.
Due to the state space explosion problem, the verification times are prohibitive for larger $N$.
We therefore, additionally, prove a general bound of $(N-1)(N-2)$ on the number of overtakes for $N \geq 4$ processes.

\paragraph{Related work.}

The mutual exclusion problem was first described around 1960.
For two processes, the main challenge is to break the tie when both processes want to access their critical section.
The first solution for this problem was Dekker's algorithm~\cite{Dij1962},
that uses a nested loop to break this tie.
The desire to eliminate this loop led to a series of algorithms~\cite{holt1978structured,DORAN1980206,Pet1981}.
Peterson's algorithm~\cite{Pet1981} is a well-known solution without nested loop.
It uses one shared variable per process to indicate the request to enter the critical section, and a single variable shared among all processes to break the tie.
Mutual exclusion algorithms for $N$ processes were first studied in~\cite{Dij1965}.
Many solutions use tournament trees~\cite{PF1977}. An alternative generalization of Peterson's algorithm uses a filter lock~\cite{Pet1981}.

Correctness of Peterson's algorithm, even for two processes, is subtle and depends on the assumptions on the environment.
For instance, it requires that shared variables are atomic, which can be deemed unrealistic as this requires a mutual exclusion algorithm in the operating system~\cite{DvGH2017}.
In other cases, fairness is required~\cite{BLW2020}, which may rule out realistic behavior~\cite{DvGH2017}.

Model checkers are commonly used to verify the correctness of mutual exclusion algorithms~\cite{GK2021,long2014modelling}.
They have also been used for performance evaluation of mutual exclusion algorithms~\cite{MS2013}, and to verify such algorithms in a timed setting~\cite{CNS2016}.
The state space explosion problem limits model checking to instances with a few processes.
Theorem provers can be used to verify arbitrary numbers of processes.
Isabelle/HOL, e.g., was used to verify Peterson's algorithm~\cite{7780229}.

\paragraph{Structure.}
The remainder of this paper is organized as follows. Section~\ref{sec:preliminaries} introduces the necessary background.
Peterson's algorithm for $N$ processes and its model and verification are described in Section~\ref{sec:peterson}.
In Section~\ref{sec:improved_algo}, the fair $N$-process mutual exclusion algorithm is described.
The paper is concluded in Section~\ref{sec:conclusion}.

\section{Preliminaries}\label{sec:preliminaries}

In this section we present the most important concepts of mCRL2 and the modal mu-calculus that we use in this paper to model and verify Peterson's algorithm.

\paragraph{mCRL2 specifications.}
The mCRL2 language~\cite{GM2014} is a process algebra with data.
It supports common data types such as Booleans $\mathbb{B}$, and numeric data types such as natural numbers $\mathbb{N}$ and their typical operations out of the box.
Additional abstract data types and operations can be defined as needed.

The behavior of a system is described using processes.
The basic building block of a process is a (parameterized) action, e.g., $a_s(42)$.
Process $x {\cdot} y$ describes the sequential composition of processes $x$ and $y$, where first $x$ and then $y$ is executed.
The choice between $x$ and $y$ is denoted using alternative composition $x + y$.
Named processes can be defined using equations, and carry parameters, e.g. $P(n \colon \mathbb{N}) = a(n) {\cdot} P(n+1)$, where $P(0)$ is the process that performs $a(0) {\cdot} a(1) {\cdot} a(2) {\cdot} \cdots$.
Within a process, the sum-operator offers a generalized alternative composition over data, e.g., $\sum_{n \colon \mathbb{N}} a_r(n)$ is equivalent to the process $a_r(0) + a_r(1) + \cdots$. A process can be restricted using the conditional operator $c \to x \mathop{\diamond}  y$, where $c$ is a Boolean expression describing the condition. The process $x$ is executed when $c$ is true, else $y$ is executed.

Processes $x$ and $y$ can be executed in parallel, using parallel composition $x \parallel y$.
For actions $a_s$ and $a_r$, $a_s \parallel a_r$ is equivalent to $a_s {\cdot} a_r + a_r {\cdot} a_s + a_s|a_r$, where $a_s|a_r$ denotes the simultaneous occurrence of $a_s$ and $a_r$.
Using communication operator $\comm{a_s|a_r \to a_c}$ we express that processes synchronize on actions $a_s$ and $a_r$, which results in action $a_c$.
This can be used to model that one process sends a value along $a_s$, and the other process receives that same value along $a_r$, resulting in handshaking communications $a_c$.
For instance $\comm{a_s|a_r \to a_c}(a_s \parallel a_r) = a_s {\cdot} a_r + a_r {\cdot} a_s + a_c$.
If we want to enforce synchronization, since we are not interested in unsuccessful communication using $a_s$ and $a_r$, we use the allow operator $\allow{a_c}$. For instance, $\allow{a_c}(\comm{a_s|a_r \to a_c}(a_s \parallel a_r)) = a_c$.

\paragraph{Modal $\mu$-calculus.}
The first order modal $\mu$-calculus is a temporal logic that extends Hennessy-Milner logic (HML)~\cite{HM1985} with fixed points and data.
Besides the standard boolean connectives $\neg$, $\land$, and $\lor$, it includes the modalities $[R]\Phi$ and $\langle R \rangle\Phi$ and the least (resp. greatest) fixed point operator $\mu$ (resp. $\nu$).
We write $\true$ inside the modalities to represent the set of all actions, so, e.g., $\langle \true \rangle \true$ holds if an action is possible from the current state, whereas $[\true]\false$ holds if no action is possible. We can specify that, from the initial state, actions other than $a$ are not possible using $[\overline{a}]\false$. $R$ represents a regular formula. For instance, property $\langle \true^* {\cdot} a \rangle \true$ holds in a state if there is a path to a state in which the action $a$ can be done. $[R]\Phi$ holds in a state if all paths characterized by $R$ end up in a state satisfying $\Phi$.

Using fixed points, we can express that action $a$ must inevitably occur within a finite number of steps using $\mu X . ([\overline{a}]X \land \langle true \rangle \true)$.

To support using data, the $\mu$-calculus also includes quantifiers $\forall$ and $\exists$.
For a detailed description see~\cite{GM2014}.

\section{Peterson's Algorithm for $N>2$ Processes}\label{sec:peterson}

Peterson's algorithm~\cite{Pet1981} can be generalized to more than two processes using a tournament tree~\cite{PF1977}.
We follow the description of the algorithm from~\cite{Fok2013}, and first introduce the tournament tree, then describe the algorithm, and finally describe the model and its verification using mCRL2~\cite{GM2014,BGK+2019}.

\subsection{Tournament trees}

Given $N$ processes that compete for the critical section, a tournament tree is a binary tree with at least $\ceil*{\frac{N}{2}}$ leaves. We identify the nodes with a number.
The root has number $0$, and all other nodes $n$ have parent $p(n) = \ceil*{\frac{n}{2}} - 1$.
Each process is associated to either the left (denoted $0$) or right ($1$) side of a node in the tree.
At any time, at most one process is associated to every side of a node.
The processes associated to internal nodes are the winners of the tournament in the left and right subtrees, respectively.
The initial position of process $i$, i.e., when it is in its non-critical section and has not yet requested access to its critical section, is on side $s = \initside(i)$ of node $n = \initnodeN(i)$, defined as follows.
\begin{align*}
	\initnodeN(i) & = 2^{\log_2 \ceil*{\frac{N}{2}}} - 1 + \floor*{\tfrac{i}{2}}\\
	\initside(i) & = i \mod 2
\end{align*}

\begin{example}
The tournament tree for $N = 3$, with processes $0 \leq i < 3$ is used as a running example throughout this paper. It is shown on the right.
\begin{floatingfigure}[r]{.5\textwidth}
	\centering
	\scriptsize
	\begin{tikzpicture}[node distance=30pt]
		\node [state,draw] (0) {$0$};
		\node [state,draw, below of= 0, left of=0,xshift=-20pt] (1) {$1$};
		\node [state,draw, below of=0, right of=0,xshift=20pt] (2) {$2$};
		\node [below left of=1] (i0) {$i = 0$};
		\node [below right of=1] (i1) {$i = 1$};
		\node [below left of=2] (i2) {$i = 2$};

		\draw[-] (0) edge node[above left] {$0$} (1)
		(0) edge node[above right] {$1$} (2);
		\draw[-,dashed] (1) edge node[left] {$0$} (i0) edge node[right] {$1$} (i1)
		(2) edge node[left] {$0$} (i2);
	\end{tikzpicture}
\end{floatingfigure}
The tournament tree has root node $0$, with left child $1$ and right child $2$. Processes $i = 0$ and $i = 1$ are initially associated with the left and right side of node $1$, respectively. Process $i = 2$ is at the left side of node $2$.
\end{example}

\subsection{Algorithm}

In Peterson's mutual exclusion algorithm for $N$ processes~\cite{PF1977}, each process that competes for access to its critical section repeatedly executes Peterson's algorithm for two processes~\cite{Pet1981} in different nodes in the tournament tree.

\begin{algorithm}[!ht]
	\KwIn{ $i,n,s,N \colon \mathbb{N}$; $i$ is the id of the process, process $i$ is currently at side $s$ of node $n$ in the tree, and $N$ is the number of processes. }

	non-critical section\;       \label{algo:petersonTT-nc}
	$\flag_{n}[s] \gets \true$\; \label{algo:petersonTT-set-flag}
	$\wait_{n} \gets s$\;        \label{algo:petersonTT-set-wait}
	\While{$\flag_{n}[1-s]$ \textbf{and} $\wait_{n} = s$}
	{$ \{ \}  $}

	\eIf{ $n = 0$	               \label{algo:petersonTT-start-if}}{
		enter the critical section\;
		leave the critical section\;
	}
	{                            \label{algo:petersonTT-recursive}
		$\mathit{Peterson}(i, \ceil{\frac{n}{2}} - 1, (n + 1) \mod 2, N)$ \;
		                           \label{algo:petersonTT-end-if}
	}

	$\flag_{n}[s] \gets \false$\;
	\If{$n = \initnodeN(t)$}{   \label{algo:petersonTT-if-init}
		$\mathit{Peterson}(i, \initnodeN(i), \initside(i),N)$}
		                          \label{algo:petersonTT-endif-init}
	\caption{$\mathit{Peterson}(i,n,s,N)$}
	\label{Algo:petersonTT}
	\end{algorithm}

If process $i$ is currently at side $s$ of node $n$ it competes with the process at side $1-s$ using Algorithm~\ref{Algo:petersonTT}.
At most one process is allowed to move up to the parent of node $n$, $p(n)$.
Peterson's algorithm in node $n$ is executed using shared variables $\flag_n[0]$, $\flag_n[1]$ to indicate whether the left (resp. right) process wants to move up in the tree. A third shared variable, $\wait_n$, is used to break the tie in case both processes want to move up.
If $\wait_n = i$ process $1-i$ has priority and is allowed to move up first, or enter the critical section in case $n=0$.

A process indicates it wants to access the critical section, and move up in the tree, by setting its flag to $\true$.
By setting $\wait_n$ to itself, it allows the other process to go first in case that process already executed lines~\ref{algo:petersonTT-nc}--\ref{algo:petersonTT-set-flag}.
If the other process has not set its flag, or arrived at line~\ref{algo:petersonTT-set-wait} last (i.e., $\wait_n = 1-i$), process $i$ can move up in the tree, and execute the algorithm again in the parent of node $n$.
If the algorithm is executed in the root and process $i$ is allowed to progress, it enters the critical section.
Upon leaving the critical section it resets all flags (in all nodes) it previously set to $\true$ back to $\false$, moving down in the tree to its initial position.
Once the flag in the initial node is set to $\false$, process $i$ repeats the algorithm, and it can request access to the critical section again if needed.


\subsection{mCRL2 Encoding}

Our model is a generalization of the model for two processes from~\cite{GK2021}, and reduces to that model when $N = 2$. Hence we only discuss the main ideas.\footnote{See \url{https://gitlab.tue.nl/y.hafidi/petersonspapermodels.git} for the full model.}

As in~\cite{GK2021}, shared variables are modelled as separate processes.
For instance, $\flag_{\node}[\side]$ is modelled as follows.
\begin{align*}
	\Flag(\node \colon \mathbb{N}, \side \colon \mathbb{N}, v \colon \mathbb{B}) & =
	\sum_{v' \colon \mathbb{B}} \mathit{set\_flag_r}(\node,\side,v') {\cdot} \Flag(v = v') \\
	& + \mathit{get\_flag_s}(\node,\side,v) {\cdot} \Flag()
\end{align*}
Here $v$ records the current value, $\mathit{set\_flag_r}$ specifies writing a new value $v'$ into the shared variable, and $\mathit{get\_flag_s}$ describes the action of sending the current value. The other shared variables are modelled analogously.

This is used in Peterson's algorithm, for process $\id$, on side $\side$ of node $\node$.
\[
\begin{array}{l}
	\mathit{Peterson}(\id \colon \mathbb{N}, \node \colon \mathbb{N}, \side \colon \mathbb{N}) = \\
	\qquad \mathit{set\_flag_s}(\node, \side, \true) {\cdot} \\
	\qquad \mathit{set\_wait_s}(\node, \side) {\cdot} \\
	\qquad ( \mathit{get\_flag_r}(\node, 1-\side, \false) + \mathit{get\_wait_r}(\node, 1-\side) ) {\cdot} \\
	\qquad ((\node = 0) \to \enter(\id) {\cdot} \leave(\id)  \\
	\qquad \phantom{(\node = 0)~~} \mathop{\diamond}  \mathit{Peterson}(\node = \ceil*{\frac{\node}{2}}-1, \side = (\node + 1) \mod 2)) {\cdot} \\
	\qquad ((\node = \initnodeN(\id)) \to \mathit{set\_flag_s}(\node, \side, \false) {\cdot} \\
	\qquad \phantom{((\node = \initnodeN(\id))) \to} \mathit{Peterson}(\node = \initnodeN(\id), \side = \initside(\id)))  \\
	\qquad \phantom{((\node = \initnodeN(\id)))} \mathop{\diamond} ~\mathit{set\_flag_s}(\node, \side, \false)
\end{array}
\]
The process follows the description in Algorithm~\ref{Algo:petersonTT}.
First, the $\flag$ and $\wait$ variables are set.
To model the busy-waiting loop, we rely on communicating actions that block as long as one of the communication partners cannot take part.
In particular, $\mathit{get\_flag_r}(\node, 1-\side, \false)$ can only be executed when also $\mathit{get\_flag_s}(\node, 1-\side, \false)$ executes, i.e., the process can only take a step if the corresponding shared variable $\flag_{\node}[\side]$ is $\false$.
If the process was executing in the root of the tournament tree, i.e., $\node = 0$, it enters and leaves the critical section.
Otherwise, it recursively executes the algorithm in the parent node in the tree.
Once the process has left the critical section, it will reset all flags to $\false$, and once it reaches its initial node, the process repeats.

To describe a particular instance of Peterson's for $N$ processes, $N$ copies of $\mathit{Peterson}$ and processes for all shared variables in each of the tree nodes are put in parallel. The communication operator $\Gamma$ is used to specify synchronization between, e.g., $\mathit{set\_flag_s}$ and $\mathit{set\_flag_r}$. Synchronization is enforced using $\nabla$.

\begin{example}
	For our running example with $N = 3$, the initialization is as follows.
	\[
	\begin{array}{l}
		\nabla_{\{\enter, \leave, \mathit{get\_flag}, \mathit{set\_flag}, \mathit{get\_wait}, \mathit{set\_wait}\}} ( \\
		\quad \Gamma_{\{\mathit{get\_flag_s}|\mathit{get\_flag_r} \rightarrow \mathit{get\_flag},
			\mathit{set\_flag_s}|\mathit{set\_flag_r} \rightarrow \mathit{set\_flag},}\\
		\quad \phantom{\Gamma_{\{}}_{\mathit{get\_wait_s}|\mathit{get\_wait_r} \rightarrow \mathit{get\_wait},
			\mathit{set\_wait_s}|\mathit{set\_wait_r} \rightarrow \mathit{set\_wait} \} } ( \\
		\quad\quad \mathit{Peterson}(0, \initnode{3}(0), \initside(0)) \parallel \\
		\quad\quad 	\mathit{Peterson}(1, \initnode{3}(1), \initside(1)) \parallel \\
		\quad\quad 	\mathit{Peterson}(2, \initnode{3}(2), \initside(2)) \parallel \\
		\quad\quad 	\Flag(0,0, \false) \parallel \Flag(0,1,\false) \parallel \mathit{Wait}(0,0) \parallel \\
		\quad\quad 	\Flag(1,0, \false) \parallel \Flag(1,1,\false) \parallel \mathit{Wait}(1,0) \parallel \\
		\quad\quad 	\Flag(2,0, \false) \parallel \Flag(2,1,\false) \parallel \mathit{Wait}(2,0)
		)
		)
	\end{array}
	\]
\end{example}

\subsection{Requirements \label{subsec:Analysis1}}

In this paper, we restrict ourselves to the verification of the three requirements described below.
These requirements are generally desirable for a mutual exclusion algorithm. It is well-known they hold for Peterson's algorithm for two processes.
They were formalized in the mu-calculus and verified in~\cite{GK2021}.

\medskip\noindent
\textbf{Mutual exclusion.}
At any time at most one process can be in the critical section.
This is formalized by saying that, invariantly, there are no two consecutive enters without an intermediate leave.
\[
[\true^* {\cdot} (\exists_{\id_1 \colon \mathbb{N}} \enter(id_1)) {\cdot} \overline{(\exists_{\id_2 \colon \mathbb{N}} \leave(id_2))}^* {\cdot} (\exists_{\id_3 \colon \mathbb{N}} \enter(id_3))]\false
\]

\medskip\noindent
\textbf{Always able to eventually request access to the critical section.}
The system should always allow processes to eventually request access to the critical section.
This is formalized as follows.
\[
[\true^*]
\forall_{\id \colon \mathbb{N}}( \id < N \implies
\langle \true^* {\cdot} \mathit{set\_flag(\initnode{N}(\id), \initside(\id),\true)} \rangle \true)
\]
Note that we capture that process $\id$ requests access to its critical section using $\mathit{set\_flag(\initnode{N}(\id), \initside(\id),\true)}$.

\medskip\noindent
\textbf{Starvation freedom.}
If a process requests access to the critical section, it is allowed to enter within a finite number of steps.
The least fixed point in the formula ensures that, after requesting access to the critical section, actions other than $\enter(\id)$ can only be taken a finite number of times, before $\enter(\id)$ is taken.
\[
\begin{array}{l}
	[\true^*] \forall_{\id \colon \mathbb{N}}( \id<N \implies\\
	\quad [\mathit{set\_flag(\initnode{N}(\id), \initside(\id),\true)}]
	(\mu X. [\overline{\enter(\id)}]X \land \langle \true \rangle \true))
\end{array}
\]

\medskip\noindent
\textbf{Bounded overtaking.}
When process $i$ requests access to the critical section, other processes enter the critical section at most $B$ times before process $i$ enters.
\[
\begin{array}{l}
	[\true^*] \forall_{\id \colon \mathbb{N}}( \id<N \implies\\
	\quad [\mathit{set\_flag(\initnode{N}(\id), \initside(\id),\true)}]
	(\nu Y(n \colon \mathbb{N} = 0) {\cdot} (n \leq B) \land \\
	\qquad [\overline{\enter(\id)}]Y(n) \land \\
	\qquad [\exists_{\id' \colon \mathbb{N}} \id' \neq \id \land \enter(\id')]Y(n+1)
\end{array}
\]
The formula says that for every process $\id$, if it requests access to the critical section, every enter action other than $\enter(\id)$ increments parameter $n$ of the greatest fixed point. If the condition $n \leq B$ is violated, then the formula is false.

\subsection{Verification \label{Subsec: verification_classical}}

We verified Peterson's algorithm for $N$ processes using the mCRL2 toolset~\cite{BGK+2019} for $N \in \{3, 4, 5\}$.

Experiments were run on a system running Ubuntu Server 18.04 LTS with 4 x Intel(R) Xeon(R) Gold 6136 CPU @ 3.00GHz (96 processors) and 3TB of RAM.
All experiments are restricted to a single processor core, with a timeout of 4 hours.

The results are shown in Table~\ref{tab:experiments-petersons-tree}. Note that the property that every process can always eventually request access to its critical section is shown in column ``Always eventually request CS''.
For each property, the table shows the time required to run the verification (in seconds), and whether the property holds (\checkmark) or not (\crossmark).

\begin{table}[!h]
	\centering
	\caption{Verification results for Peterson's algorithm for $N$ processes.
	\label{tab:experiments-petersons-tree}}

	\begin{tabular}{@{}rcrccrccrcc@{}}

			\toprule
			\multirow{2}{*}{$N$}    & & \multicolumn{2}{c} {Mutual exclusion} & & \multicolumn{2}{c}{\makecell{Always eventually request CS}} & &  \multicolumn{2}{c}{Starvation freedom}    \\
			\cmidrule{3-4}
			\cmidrule{6-7}
			\cmidrule{9-10}
			& & \multicolumn{1}{c}{Time}   & Result      & & \multicolumn{1}{c}{Time}   & Result & & \multicolumn{1}{c}{Time}   & Result \\
			\midrule
			3   & & ~~~0.34 & \checkmark  & & ~~~0.69 & \checkmark & & ~~~0.37 & \crossmark    \\
			4   & & ~~~1.41 & \checkmark  & & ~~~5.70 & \checkmark & & ~~0.93 & \crossmark     \\
			5   & & 139.76 & \checkmark   & & ~~~775.90 & \checkmark & & 34.74 & \crossmark    \\
			\bottomrule
	\end{tabular}
	\end{table}

For all values of $N$, the algorithm satisfies mutual exclusion and each process is always able to eventually request access to the critical section. However, starvation freedom (and hence also bounded overtaking) is not satisfied. We explain why this is the case using our running example.

\begin{example}
We show the counterexample to starvation freedom for $N = 3$ by listing steps executed by each of the processes in Figure~\ref{fig:counterexample}.
The counterexample shows that process 0 requests access to the critical section ($\flag_1[0] := \true$), but never enters.
Instead, process 2 accesses its critical section infinitely many times, by setting its flag in node 2 immediately after finishing the execution of the algorithm.
Until process 0 sets $\flag_0[0]$ to $\true$, it can always be overtaken by process 2.
\end{example}

\begin{figure}[h]
	\centering
	\begin{tabular}{lll}
		$\id = 0$             & $\id = 1$             & $\id = 2$ \\ \hline
		$\flag_1[0] := \true$ & & \\
		$\wait_1 := 0$        & & \\
		$\flag_1[1] = \false$ & & \\
		& $\flag_1[1] := \true$ & \\
		& $\wait_1 := 1$        & \\
		&                       & $\flag_2[0] := \true$ \\
		&                       & $\wait_2 := 0$ \\
		&                       & $\flag_2[1] = \false$ \\
		&                       & $\flag_0[1] := \true$ \\
		&                       & $\wait_0 := 1$ \\
		&                       & $\flag_0[0] = \false$ \\						&                       & enter CS \\
		&                       & leave CS \\
		&                       & $\flag_0[1] := \false$ \\
		&                       & $\flag_2[0] := \false$ \\
		&                       & /* repeat */
	\end{tabular}
	\caption{Counterexample for starvation freedom ($N=3$)}
	\label{fig:counterexample}
\end{figure}

One could argue this counterexample is unfair. In particular, process 0 continuously wants to set $\flag_0[0]$ to $\true$, but never gets the chance to do so because process $2$ always takes priority.
Similar counterexamples exist where process $0$ continuously wants to read the flag of side $1$ in node $1$, one of the processes $0$ or $1$ continuously wants to set $\wait_1$, or wants to read $\wait_1$.\footnote{Recall that node $1$ is $\initnode{3}(0)$.}

\medskip\noindent
\textbf{Starvation freedom under fairness.}
We modify the formula for starvation freedom in such a way that unfair paths satisfy the property.
For the sake of simplicity, we only give the property for process 0, in the case where $N = 3$.
\[
\begin{array}{l}
	[\true^*] [\mathit{set\_flag(\initnode{3}(0), \initside(0),\true)}]\\
	\quad ( \nu X. \mu Y. \\
	\hspace{1cm} (\langle  \exists_{\side \colon \mathbb{N}}  set\_wait(\initnode{3}(0),\side)  \lor \\
	\hspace{1.2cm} \mathit{set\_flag(0,0, \true)} \lor \\
	\hspace{1.2cm} \mathit{get\_flag(\initnode{3}(0),1 -\initside(0), \false)}
	\lor \\
	\hspace{1.2cm} \mathit{set\_flag(\initnode{3}(0), 1 -\initside(0),\false)} \lor \\
	\hspace{1.2cm} \exists_{\side \colon \mathbb{N}}  \mathit{get\_wait(\initnode{3}(0),side)}
	\rangle \true \\
	\hspace{1cm} \land [\overline{\enter(0)}]X) \\
	\quad \lor
	(\langle \true \rangle \true \land [\overline{ \enter(0)}]Y))
\end{array}
\]
This formula says that whenever process 0 requests access to the critical section, each infinite execution without an $\enter(0)$ action ends in an infinite sequence of states in which one of the actions indicated previously is continuously enabled.
This property holds for the case where $N = 3$.
However, this property is not ideal. When adding more layers to the tournament tree, the property has to be adapted to take fairness on the additional levels into account as well.

Note that we do not consider all actions of process 0 in the formula.
This is because other actions, such as $\mathit{set\_wait}(0,0)$, $\mathit{set\_flag}(0,0,\false)$, $\enter(1)$ and $ \leave(1)$ affect the global progress, hence their execution will automatically be enforced by the algorithm  at some point, and no fairness assumption is needed on them.

Indeed, if we assume weak fairness, the algorithm is starvation free, but does not satisfy bounded overtaking.
However, we also recall that Peterson's algorithm for two processes is starvation free without requiring fairness assumptions. The reason for this is that mutual dependency between the two processes based on their $\wait$ variables forces progress. 
This observation inspires our fair algorithm in the next section.

\section{Fair $N$-process Mutual Exclusion Algorithm}\label{sec:improved_algo}

We modify Peterson's algorithm for $N$ processes such that any process that requests access to the critical section is eventually forced into its critical section by all other processes. The main idea is that, each time when a process leaves its critical section, it will wait for another process, $t$, in a cyclic way.

We first introduce some notation.
The sibling of process $i$ ($0 \leq i < N$) is the process $j$ that shares the same initial node in the tournament tree if such process exists, and it is undefined otherwise. 
\[
\siblingN(i) = \begin{cases}
	j & \text{if $0 \leq j < N$ s.t. $j \neq i$ and $\initnodeN(j) = \initnodeN(i)$} \\
	\bot & \text{if for all $0 \leq j < N$ s.t. $j \neq i$, $\initnodeN(j) \neq \initnodeN(i)$}
\end{cases}
\]

If process $i$ is waiting on process $t$, it will next wait for process $\nextN(t,i)$.
\[
\nextnode{N}(t, i) = \begin{cases}
	(t + 1) \mod N & \text{if $\initnodeN((t + 1) \mod N) \neq$} \\ & \text{\phantom{if }$\initnodeN(i)$}  \\
	\nextN((t + 1) \mod N, i) & \text{otherwise}
\end{cases}
\]
The sibling of $i$ is skipped, as it will be forced to progress based on the properties of the two-process Peterson's algorithm executed in the shared initial node.

\begin{algorithm}[!ht]
	\KwIn{ $i,n,s,N,t \colon \mathbb{N}$; (\ldots) $t$ is the process on which $i$ waits after it has left the critical section. }
	\setcounter{AlgoLine}{12}
	$\vdots$\\
	\If{$n = \initnodeN(t)$}{\label{alg:wait_t}
		\While{$\flag_{\initnodeN(t)}\left[\initside(t)\right]$}
		{$ \{ \}  $}
		$\mathit{PetersonFair}(i, \initnodeN(i), \initside(i), N, \nextN(t,i))$}
		\caption{$\mathit{PetersonFair}(i,n,s,N,t)$}
	\label{Algo:petersonTTImproved}
\end{algorithm}

The modifications required to make the algorithm fair are shown in Algorithm~\ref{Algo:petersonTTImproved}. An additional parameter $t$ is added to keep track of the process on which the algorithm will wait.
The recursive call on line~\ref{algo:petersonTT-recursive} passes this parameter on unchanged.
Lines~\ref{algo:petersonTT-if-init}--\ref{algo:petersonTT-endif-init} are changed such that, upon leaving its critical section, process $i$ blocks waiting for process $t$ as long as process $t$ has set its flag in its initial node in the tournament tree.
Once the flag becomes false, process $i$ is allowed to proceed as usual, while waiting for $\nextN(t,i)$ after it leaves the critical section the next time.
If process $t$ has set its flag, and has not yet entered the critical section, eventually the algorithm will end up in a situation where all processes are waiting for $t$, and $t$ is forced to progress into its critical section.

\subsection{mCRL2 Encoding}

The mCRL2 model of the improved algorithm is mostly analogous to the model of Algorithm~\ref{Algo:petersonTT}.
Parameter $t$ is added. Other than that, only the last conditional in the process needs to be changed. Once process $\id$ has set the flag to $\false$ in its initial node upon leaving the critical section, it blocks until the flag of process $t$ in the corresponding initial node is $\false$. We use the same construct for modelling a busy-waiting loop as before. The resulting change is as follows.
\[
\begin{array}{l}
	\mathit{PetersonFair}(\id \colon \mathbb{N}, \node \colon \mathbb{N}, \side \colon \mathbb{N}, t \colon \mathbb{N}) = \\
	\hspace{0.7cm}
	$\vdots$\\

	\hspace{0.7cm} ((\node = \initnode{N}(\id)) \\
	\hspace{1.4cm} \to (\mathit{set\_flag\_s}(\node, \side, \false) {\cdot} \\
	\hspace{1.4cm} \phantom{\to (} \mathit{get\_flag\_r}(\initnode{N}(t), \initside(t), \false) {\cdot} \\
	\hspace{1.4cm} \phantom{\to (} \mathit{PetersonFair}(\node = \initnode{N}(\id), \side = \initside(\id), t = \nextN(t))) \\
	\hspace{1.45cm} \mathop{\diamond}  \mathit{set\_flag\_s}(\node, \side, \false) )

\end{array}
\]

\subsection{Verification}

Using the same setup as in Section~\ref{Subsec: verification_classical} we have verified our fair algorithm. The results are shown in Table~\ref{tab:verification-results-fair-algorithm}, where '$\texttt{to}$' denotes a timeout. We include the results for bounded overtaking, where $B$ is the lowest bound for which the property holds.

	\begin{table}
	\centering
	\caption{Verification results for the fair mutual exclusion algorithm for $N$ processes.}
	\label{tab:verification-results-fair-algorithm}
	\begin{tabular}{@{}rcrccrccrccrccr@{}}
		\toprule
		\multirow{2}{*}{$N$}    & & \multicolumn{2}{c} {Mutual exclusion}  & & \multicolumn{2}{c}{\makecell{Always eventually \\ request CS}} & & \multicolumn{2}{c}{Starvation freedom} & & \multicolumn{3}{c} {Bounded overtaking}  \\
		\cmidrule{3-4}
		\cmidrule{6-7}
		\cmidrule{9-10}
		\cmidrule{12-14}
		& & \multicolumn{1}{c}{Time}   & Result      & & \multicolumn{1}{c}{Time}   & Result & & \multicolumn{1}{c}{Time}   & Result  & & \multicolumn{1}{c}{Time}   & Result & $B$ \\
		\midrule
		3   & & ~~~0.82 & \checkmark  & & ~~~2.03 & \checkmark   & & ~~~1.94 & \checkmark & & ~~~2.70 & \checkmark & 4   \\
		4   & & ~~~42.03 & \checkmark   & & ~~~199.18 & \checkmark & & ~~164.01 & \checkmark & & ~352.82 & \checkmark & 6    \\
		5   & & $\texttt{to}$ & N/A  & & $\texttt{to}$ & N/A & & $\texttt{to}$ & N/A & & $\texttt{to}$ & N/A & 12    \\
		\bottomrule
	\end{tabular}
\end{table}

From Table~\ref{tab:verification-results-fair-algorithm}
we can observe that all of the properties are satisfied for all values of $N$ that we verified.
Furthermore, the verification time grows rapidly, in fact, much more so than for the original algorithm. This is due to the different combinations of processes on which each process can be waiting.
For $N=5$ verification times out, meaning it already requires over 4 hours.

\subsection{Analysis}

Due to the state space explosion problem, we are unable to verify correctness of the algorithm for arbitrary values of $N$ using model checking.
That Peterson's algorithm for $N$ processes satisfies mutual exclusion is well-known. As the guarding mechanism for the critical section remains unaltered in our proposed algorithm, mutual exclusion is preserved.
We therefore give a manual proof of an upper bound on the number of times a process can be overtaken. Starvation freedom follows immediately from this.

To determine a bound, we first characterize $\overtakesN(j)$, which is the number of times a process $j$ can be overtaken by any other process, once it has requested access to the critical section.
The number of times process $i$ can overtake $j$, denoted $\overtakeN(i,j)$, is bounded by the number of different values parameter $t$ of $i$ can have until the value of $t$ becomes $j$, since at that point process $i$ is blocked until $j$ has left (and thus entered) its critical section.
This is specified as follows.

\begin{definition}\label{def:overtakes}
	Let $N$ be the number of processes, and $0 \leq i,j < N$. Then
	\begin{align*}
		\overtakesN(j) & = \sum_{i = 0}^{N-1} \overtakeN(i, j) \\
		\overtakeN(i,j) & = \begin{cases}
		0 & \text{if $i = j$}\\
		2 & \text{if $i = \siblingN(j)$} \\
		\left| \{ \nextN(t,i) \mid 0 \leq t < N \} \right| & \text{otherwise}
	\end{cases}
	\end{align*}
\end{definition}

\begin{toappendix}
The following lemma expresses that process $i$ can wait for any process that is not $i$ itself, and that is not a sibling of $i$. These processes are calculated using $\nextN$.
\begin{lemmarep}\label{lem:next-images}
	Let $N > 2$, and $0 \leq i < N$. Then we have the following:
	\[
	\{ \nextN(t, i) \mid 0 \leq t < N \} = \{ j \mid 0 \leq j < N, j \neq i, j \neq \sibling{N}(i) \}
	\]
\end{lemmarep}
\begin{proof}
	Fix $N > 2$ and $0 \leq i < N$. We prove set inclusion in both directions separately.
	\begin{itemize}
		\item[$\subseteq$] We have to show that $\{ \nextN(t, i) \mid 0 \leq t < N \} \subseteq \{ j \mid 0 \leq j < N, j \neq i, j \neq \sibling{N}(i) \}$.
		Fix $x \in \{ \nextN(t, i) \mid 0 \leq t < N \}$. So, there exists $t$, $0 \leq t < N$ such that $x = \nextN(t,i)$. Let $t$ be such.
		We have to show that $x \in \{ j \mid 0 \leq j < N, j \neq i, j \neq \sibling{N}(i) \}$.

		First, observe that any value returned by $\nextN$ is of the form $(t+1) \mod N$, hence $0 \leq x = \nextN(t,i) < N$.
		Now, we prove that $x \neq i$.
		Towards a contradiction, suppose that $x = i$, then $x = \nextN(t, i) = (t + 1) \mod N$, hence, $\initnodeN(i) \neq \initnodeN((t+1) \mod N)$, but, as $i = (t+1) \mod N$, $\initnodeN((t+1) \mod N) = \initnodeN(i)$, contradiction.
		Finally, we prove that $x \neq \siblingN(i)$.
		Again, towards a contradiction, suppose that $x = \siblingN(i)$, then $x = \nextN(t, i) = (t + 1) \mod N$, so $\siblingN(i) = \nextN(t, i) = (t + 1) \mod N$. Hence $\initnode(i) \neq \initnodeN((t+1) \mod N)$, but, as $\siblingN(i) = (t + 1) \mod N$, $\initnodeN((t+1) \mod N) = \initnodeN(i)$, which is a contradiction.

		So, in all cases where $\nextN$ returns a value $x$, we have shown that $0 \leq x < N$ and $x \neq i$ and $x \neq \siblingN(i)$.
		Note that $\nextN$ is well-defined, as $N > 2$, and there are only two values of $t$ such that $\initnodeN((t+1) \mod N) = \initnodeN(i)$.

		\item[$\supseteq$] We have to prove that $\{ j \mid 0 \leq j < N, j \neq i, j \neq \sibling{N}(i) \} \subseteq	\{ \nextN(t, i) \mid 0 \leq t < N \}$.
		Fix $x \in \{ j \mid 0 \leq j < N, j \neq i, j \neq \sibling{N}(i) \}$, i.e., $0 \leq x < N$, and $x \neq i$ and $x \neq \sibling{N}(i)$. We have to prove that $x \in \{ \nextN(t, i) \mid 0 \leq t < N \}$. So, we need to prove there exists a $t$, $0 \leq t < N$, such that $\nextN(t,i) = x$.
		Choose $t = (x - 1) \mod N$. Clearly $0 \leq (x - 1) \mod N < N$.
		Note that $(((x - 1) \mod N) + 1) \mod N = x$.
		As $x \neq i$ and $x \neq \siblingN(i)$, $\initnodeN(i) \neq \initnodeN(x)$ we are in the first case of the definition of $\nextN$. Therefore, $\nextN((x - 1) \mod N), i) = ((x - 1) \mod N) + 1) \mod N = x $. \qedhere

	\end{itemize}
\end{proof}

Using this lemma, we can now establish the following relation between the number of images of $\nextN$, and the number of processes, $N$.
\begin{lemmarep}\label{lem:size-of-next-image}
	Let $N > 2$, and $0 \leq i < N$. Then $\left| \{ \nextN(t,i) \mid 0 \leq t < N \} \right| = N - 1$ if $\siblingN(i) = \bot$, and $N-2$ otherwise.
\end{lemmarep}
\begin{proof}
	Fix $N > 2$ and $0 \leq i < N$. We prove both cases separately.
	\begin{enumerate}
		\item Suppose $\siblingN(i) = \bot$. We have to show that $\left| \{ \nextN(t,i) \mid 0 \leq t < N \} \right| = N - 1$.
		Observe that, as $\siblingN(i) = \bot$, according to Lemma~\ref{lem:next-images}, $\{ \nextN(t,i) \mid 0 \leq t < N \} = \{ j \mid 0 \leq j < N, j \neq i \}$, hence it immediately follows that $\left| \{ \nextN(t,i) \mid 0 \leq t < N \} \right| = \left|  \{ j \mid 0 \leq j < N, j \neq i \} \right| = N - 1$.
		\item Suppose $\siblingN(i) = j$ for $0 \leq j < N$ and $j \neq i$.
		We have to show that $\left| \{ \nextN(t,i) \mid 0 \leq t < N \} \right| = N - 2$.
		According to Lemma~\ref{lem:next-images}, $\{ \nextN(t,i) \mid 0 \leq t < N \} = \{ k \mid 0 \leq k < N, k \neq i, k \neq \siblingN(i) \}$. As $\siblingN(i) = j$, this simplifies to $\{ k \mid 0 \leq k < N, k \neq i, k \neq j \}$. Since $j \neq i$, it immediately follows that $\left| \{ \nextN(t,i) \mid 0 \leq t < N \} \right| = \left|  \{ k \mid 0 \leq k < N, k \neq i, k \neq j \} \right| = N - 2$. \qedhere
	\end{enumerate}
\end{proof}
\end{toappendix}

The key observation now is that $\overtakesN(j)$ characterizes the number of times process $j$ can be overtaken in Algorithm~\ref{Algo:petersonTTImproved}. This follows from the structure of the algorithm and the definition of $\nextN$, and is formalized as follows.

\begin{lemmarep}\label{lem:overtake-alg}
	Let $N$ be the number of processes. If process $j$ requests access to its critical section, Algorithm~\ref{Algo:petersonTTImproved} guarantees that other processes enter their critical section at most $\overtakesN(j)$ times before process $i$ enters its critical section.
\end{lemmarep}
\begin{appendixproof}
	Let $N$ and $j$ be arbitrary. Given the definition of $\overtakesN$, it suffices to consider the number of times each process $i$ can overtake process $j$ before $j$ enters its critical section. Fix $i$. We distinguish the following cases:
	\begin{itemize}
		\item $i = j$. Then trivially, process $i$ cannot enter the critical section before process $j$, so $\overtakeN(i,j) = 0$ is consistent with the algorithm.
		\item $i = \siblingN(j)$, then process $i$ and $j$ are in the same initial node in the tournament tree.
		It therefore follows from the bounded overtaking property of Peterson's algorithm for two processes that process $i$ can proceed to the next level twice before process $j$ is forced to proceed.
		Hence, process $i$ can also continue to the critical section twice before process $j$ enters its critical section. So, $\overtakeN(i,j) = 2$ is consistent with the algorithm.
		\item $i \neq j$ and $i \neq \siblingN(j)$. Each time process $i$ leaves its critical section, it will wait for some process $t$ on line~\ref{alg:wait_t}; it iterates through all processes that are not a sibling of $j$, until $t = i$. When $t = i$, and process $i$ is still waiting to enter the critical section, the algorithm blocks until process $i$ has left (and therefore entered) the critical section.
		If $j$ has a sibling, the number of processes that $j$ can iterate through before being blocked waiting for $i$ to enter its critical section therefore is $N-2$ if $j$ has a sibling, and $N-1$ otherwise. This is consistent with $\left| \{ \nextN(t,i) \mid 0 \leq t < N \} \right|$ according to Lemma~\ref{lem:size-of-next-image}. \qedhere
	\end{itemize}

\end{appendixproof}
Now that we have established a link between the algorithm and $\overtakesN$, we can focus on the latter definition to determine the bound on the number of overtakes.
Following the case distinction in Definition~\ref{def:overtakes}, the exact number of times process $i$ can be overtaken depends on whether $i$ has a sibling, and whether all processes have a sibling.
This results in the following property.

\begin{lemmarep}\label{lem:overtakes}
	\label{lem:overtakes_no_sibling}
	\label{lem:overtakes_sibling}
	Let $N > 2$ and $0 \leq i < N$. Then
	\[
	\overtakesN(i) = \begin{cases}
		2 + (N - 2)^2 & \text{if $\siblingN(i) \neq \bot$ and $N$ is even}\\
		3 + (N - 2)^2 & \text{if $\siblingN(i) \neq \bot$ and $N$ is odd} \\
		(N - 1)(N - 2) & \text{otherwise (i.e. $\siblingN(i) = \bot$)}
	\end{cases}
	\]
\end{lemmarep}
\begin{appendixproof}
	Fix $N > 2$ and let $i$ be such that $0 \leq i < N$. We distinguish cases on whether $\siblingN(i)$ is defined.
	\begin{itemize}
		\item $\siblingN(i) \neq \bot$. We calculate as follows.
		\begin{align*}
			\overtakesN(i) & = \sum_{j = 0}^{N-1} \overtakeN(j,i) \\
			& = 0 + 2 + \sum_{j = 0, j \neq i, j \neq \siblingN(i)}^{N-1} \overtakeN(j, i) \\
			& = 2 + \sum_{j = 0, j \neq i, j \neq \siblingN(i)}^{N-1} \left| \{ \nextN(t,j) \mid 0 \leq t < N \} \right|
		\end{align*}
		We know that $\siblingN$ is defined for all processes in case $N$ is even, and it is defined for all processes except $N - 1$ in case $N$ is odd. We therefore distinguish two cases.
		\begin{itemize}
			\item $N$ is even. Then according to Lemma~\ref{lem:size-of-next-image}, $\left| \{ \nextN(t,j) \mid 0 \leq t < N \} \right| = N - 2$ for all $j$. So,
			\begin{align*}
				& \phantom{=} 2 + \sum_{j = 0, j \neq i, j \neq \siblingN(i)}^{N-1} \left| \{ \nextN(t,j) \mid 0 \leq t < N \} \right| \\
				& = 2 + \sum_{j = 0, j \neq i, j \neq \siblingN(i)}^{N-1} (N - 2) \\
				& = 2 + (N - 2)^2.
			\end{align*}
			\item $N$ is odd. Then $\siblingN$ is defined for all processes, except process $N - 1$, so according to Lemma~\ref{lem:size-of-next-image}, $\left| \{ \nextN(t,j) \mid 0 \leq t < N \} \right| = N - 2$ for all $0 \leq j < N-1$, and $\left| \{ \nextN(t,N-1) \mid 0 \leq t < N \} \right| = N - 1$. So,
			\begin{align*}
				& \phantom{=} 2 + \sum_{j = 0, j \neq i, j \neq \siblingN(i)}^{N-1} \left| \{ \nextN(t,j) \mid 0 \leq t < N \} \right| \\
				& = 2 + \left( \sum_{j = 0, j \neq i, j \neq \siblingN(i)}^{N-2} (N - 2)\right)  + (N - 1) \\
				& = 2 + (N - 3)(N - 2) + (N - 1) \\
				& = 2 + (N - 3)(N - 2) + (N - 2) + 1 \\
				& = 3 + (N - 2)^2.
			\end{align*}
		\end{itemize}

		\item $\siblingN(i) = \bot$.
		Observe that, according to the definition of the tournament tree, as $\siblingN(i) = \bot$, it must be the case that $N$ is odd, and $i = N - 1$.
		We now calculate as follows.
		\begin{align*}
			\overtakesN(i) & = \overtakesN(N - 1) \\
			& = \sum_{i = 0}^{N-1} \overtakeN(i, N-1) \\
			& = \sum_{i = 0}^{N - 2} \left| \{ \nextN(t,i) \mid 0 \leq t < N \} \right|
		\end{align*}
		Note that at the last step we use that, as $\siblingN(N-1) = \bot$, there is no process $i$ such that $\siblingN(i) = N - 1$, and $\overtakesN(N-1,N-1) = 0$.
		As for all $0 \leq i < N - 1$, $\siblingN(i)$ is defined, we have that
		$\sum_{i = 0}^{N - 2} \left| \{ \nextN(t,i) \mid 0 \leq t < N \} \right| = \sum_{i = 0}^{N - 2} (N - 2)$, according to Lemma~\ref{lem:size-of-next-image}.
		Finally, $\sum_{i = 0}^{N - 2} (N - 2) = (N-1)(N-2)$, hence $\overtakesN(i) = (N-1)(N-2)$. \qedhere
	\end{itemize}
\end{appendixproof}

Using this lemma, it is not hard to see that for all $i$, $\overtakesN(0) \geq \overtakesN(i)$ if $N = 3$, and $\overtakesN(N-1) \geq \overtakesN(i)$ otherwise.
Hence, to determine the bound for a given $N$, we either need to consider the number of times process $0$ or process $N - 1$ gets overtaken, depending on $N$.

\begin{toappendix}
\begin{propositionrep}\label{prop:overtakes_max}
	\mbox{}
	\begin{itemize}
		\item If $N = 3$, then for all $i$ s.t. $0 \leq i < N$, $\overtakesN(0) \geq \overtakesN(i)$.
		\item If $N > 3$, then for all $i$ s.t. $0 \leq i < N$, $\overtakesN(N - 1) \geq \overtakesN(i)$.
	\end{itemize}
\end{propositionrep}
\begin{proof}
	We prove both cases separately.
	First, assume that $N = 3$. Observe that $\overtakesN(0) = \overtakesN(1) = 3 + (N - 2)^2 = 4$ according to Lemma~\ref{lem:overtakes_sibling}. Furthermore, $\siblingN(2) = \bot$, hence according to Lemma~\ref{lem:overtakes_no_sibling}, $\overtakesN(2) = (N-1)(N-2) = 2$, so clearly the result follows.

	Now, assume that $N > 3$. We distinguish cases on the parity of $N$.
	\begin{itemize}
		\item $N$ is even. Then according to Lemma~\ref{lem:overtakes_sibling}, $\overtakesN(i) = 2 + (N - 2)^2$ for all $i$, and the result follows immediately.
		\item $N$ is odd. Fix arbitrary $i$. The case where $i = N - 1$ is trivial, so assume that $i < N - 1$. We have to show that $\overtakesN(N-1) \geq \overtakesN(i)$.
		We reason as follows.
		\begin{align*}
			\overtakesN(N - 1) &= (N - 1)(N - 2) \\
			& = (N - 2) + (N - 2)^2 \\
			& \geq^{\dagger} 3 + (N - 2)^2 \\
			& = \overtakesN(i)
		\end{align*}
		Where at $^\dagger$ we have used that $N - 2 \geq 3$ as from the assumptions that $N > 3$ and $N$ is odd it follows that $N \geq 5$.
		Furthermore, we have used Lemma~\ref{lem:overtakes_no_sibling}. \qedhere
	\end{itemize}
\end{proof}
\end{toappendix}

From this observation and Lemmas~\ref{lem:overtake-alg} and~\ref{lem:overtakes} we get the following bound.
\begin{theoremrep}\label{theorem:boundedOvertaking}
	Let $N \geq 4$ be the number of processes. If a process requests access to its critical section, then other processes enter their critical section at most $(N-1)(N-2)$ times before this process enters its critical section.
\end{theoremrep}
\begin{appendixproof}
	Let $N$ be the number of processes, and assume $N \geq 4$, then according to Proposition~\ref{prop:overtakes_max}, the maximum number of overtakes is determined by process $N-1$. We distinguish two cases:
	\begin{itemize}
		\item $N$ is even. Then $\siblingN(N-1) \neq \bot$, and $\overtakesN(N-1) = 2 + (N-2)^2 \leq (N-1)(N-2)$ according to Lemma~\ref{lem:overtakes}.
		\item $N$ is odd. Then $\siblingN(N-1) = \bot$, hence according to Lemma~\ref{lem:overtakes},
		\begin{align*}
		\overtakesN(N-1) &= (N-1)(N-2) \\
		&= (N-2) + (N-2)^2 \\
		&= (N-1)(N-2). \qedhere
	    \end{align*}
	\end{itemize}
\end{appendixproof}
So, the algorithm introduced in this section satisfies bounded overtaking. Using Lemma~\ref{lem:overtakes}, in case $N=3$, the bound is 4.

\section{Conclusion}\label{sec:conclusion}

We studied the generalization of Peterson's algorithm to $N$ processes using a tournament tree.
Using the mCRL2 language and toolset we showed that the algorithm is not starvation free if no additional fairness assumptions are imposed.
We propose a fair mutual exclusion algorithm for $N$ processes.
Using the mCRL2 model checker we verified that for small $N$ the algorithm is starvation free, and, in fact, satisfies bounded overtaking.
For the general case, when $N \geq 4$, we additionally present a proof that the bound is $(N-1)(N-2)$.

\subsubsection*{Acknowledgements}
This work was supported partially by the MACHINAIDE project (ITEA3, No. 18030).
\bibliographystyle{splncs04}
\bibliography{bib_peterson}

\end{document}